# Identification and characterization of dominant microflora isolated from selected ripened cheese varieties produced in Uganda

Andrew Mwebesa Muhame[1*]; Ediriisa Mugampoza[1]; Paul Alex Wacoo[2]; Stellah Byakika[3]

[1] *Department of Food Science and Technology, Kyambogo University, P.O BOX 1 Kyambogo*

[2] *Department of Medical Biochemistry, Makerere University P.O BOX 7062 Makerere*

[3] *Department of Food Technology and Nutrition, Makerere University P.O BOX 7062 Makerere*

*Correspondence should be addressed to Andrew Mwebesa Muhame; amuhame@kyu.ac.ug

## Abstract

In this study, the predominant lactic acid bacteria (LAB) isolates were obtained from Gouda, Jack, Cheddar, and Parmesan cheeses produced in Uganda. The isolates were identified through Gram staining, catalase and oxidase tests, and 16S rDNA sequencing. Approximately 90% of the isolates were cocci (n=192), including *Streptococcus, Enterococcus*, and *Lactococcus*. The remaining 10% were identified as rod-shaped bacteria, primarily belonging to the Lactobacillus species (n=23). BLAST analysis revealed that *Pediococcus pentosaceus* dominated in all cheese samples (23.7%, of the total 114 isolates). This was followed by uncultured bacterium (15.8%), uncultured Pediococcus species (13.2%), *Lacticaseibacillus rhamnosus* (8.8%) among others

*Key words*: Lactic acid bacteria, Cheese,

___________________________________________________________________________

## 1. Introduction

Cheese is a fermented milk product that dates back to Neolithic times (Mayo et al., 2021). Milk from various species (including cows, goats, sheep, donkeys, buffaloes, and dromedary camels), as well as their mixtures, is used along with different technological processes such as coagulation, curd cutting, whey drainage, washing, heating, pressing, immersion, wrapping, and the addition of spices to create a diverse range of cheeses. This makes cheese one of the most diverse of all foodstuffs (Fox et al., 2017). It varies in the sensory properties depending on the milk type used, the feed given to the distinct species of animals, the manufacturing processes involved, the ripening environment, period of ripening, the microorganisms involved in terms of type, numbers and their activity (Bittante et al., 2022). This offers distinct textures and flavors of cheese (Patel et al., 2024).

There are various types of microorganisms have been isolated from cheeses for example *Streptococcus, Lactococcus, Lactobacillus, Pediococcus, Leuconostoc* and many pathogens from Urfa cheeses ( Uraz et al. (2008). LAB play various roles among which they are vital probiotics in the food processing industry (Qi et al., 2021). Cheese is processed mostly in an open environment, a variety of microorganisms from the environment contribute to the fermentation and post- ripening of the cheese (Zhang et al., 2020). The characterization of the indigenous LAB microflora in locally processed ripened cheese was carried out with the purpose of developing specific starter cultures, which would allow the making of a safe and uniform product from pasteurized milk.

The adjunct and surface ripening starter cultures marketed today for smear cheeses are inadequate for mimicking the real diversity encountered in cheese microbiota (Irlinger et al., 2021).

The surface-ripening cultures marketed today are far from adequate for dealing with the diversity encountered in cheese microbiota and managing processes to support the growth of desirable microorganisms whilst preventing the undesirable ones. That is why understanding cheese microbial ecology is crucial to produce a product with a constant level of quality and safety.

## 2. Materials and Methods

### 2.1. Sample origin and preparation

Four different varieties of ripened cheese were obtained in triplicate from the factory located in Mbarara city, Uganda, including three 3-month-aged Gouda cheeses from the Netherlands as reference samples. The Ugandan varieties selected for the study were Gouda, Parmesan, Cheddar, and Jack cheese, all produced by Sanatos Dairies (U) Limited. Each cheese sample was approximately three months old. All samples were transferred aseptically to Makerere University Food Science microbiology laboratory and analyzed within 8 h of collection. Each sample (25 g) was weighed into a stomacher bag, diluted with 225 ml of sterile peptone water and homogenized using a stomacher machine (specs) at 230 rpm for 2 min. This sample was considered as the $10^{-1}$ dilution. Further 10-fold dilutions were prepared in sterile peptone water up to $10^{-7}$ (you may have to cite the author of this prep method).

## 2.2 Microbial Analyses

Each of the diluted samples (0.1 ml) was inoculated in triplicate on the various media by spread plating. All plates were incubated under appropriate conditions of temperature and aeration. After incubation, plates with colonies in the range 30-300 were enumerated using a colony counter (model OLS26, Ser No. Q71839002, UK) Stuart Scientific, England) and used to establish microbial counts from the different samples using the following formula and expressed as log cfu/g.

Colony forming units/g = $\frac{\text{(count x dilution factor)}}{\text{sample volume plated (ml)}}$

Table 1. Incubation conditions for various inoculated culture media

| Medium | Temperature (°C) | Time | Environment | Target |
|---|---|---|---|---|
| Nutrient agar | 30 | 48 h | Aerobic | Total plate counts |
| M17 agar | 30 | 48 h | Aerobic | Lactococci |
| MRS agar | 30 | 48 h | Anaerobic | Total LAB |

## 2.3. Microbial isolation, purification and presumptive identification

After enumeration of viable counts, the plates containing 30-300 colonies were used for obtaining of microbial isolates that were used for further identification and characterization. The colonies with different morphologies (color, size, texture, etc.) were purified by streaking twice on the respective selective media agar and appropriately incubated as in Table 1. One colony from each of the purified cultures was inoculated in 10 ml of respective selective media broths and appropriately incubated. A 1 ml culture containing 20% sterile glycerol was stored in a sterile 1.5 ml Eppendorf tube at -25°C until further examination for cell morphologies, Gram reaction and biochemical characteristics using conventional microbiology methods for presumptive identification. Gram staining, catalase and oxidase tests were performed following the protocol described by Mugampoza et al. (2020).

## 2.4 Biochemical Analyses
### 2.4.1 Gram staining

With the help of a dropper, one drop of peptone water was added on a glass slide into which a microbial colony was dispersed to produce a smear. The smear was air dried at room temperature for 10 min and heat fixed by passing the slide through a Bunsen flame 3-5 times before staining with crystal violet for 1 min. Crystal violet was washed off with water followed by addition of Lugol's iodine for 30 s. Excess iodine was washed off using tap water and then 96% ethanol was applied for 1 min while washing off the excess with tap water. The smear was eventually counter stained with safranin for 30 s. Excess safranin was washed off with tap water and the slide allowed to air-dry for 10 min at room temperature. A light microscope (Olympus U-RFL-T, BX51, GmbH, Hamburg, Germany) at a magnification of X1000 was used to examine the Gram reaction, and cell shapes and arrangement for the different microbial isolates. *Escherichia coli* and *Staphylococcus aureus* were used as the Gram-negative and Gram-positive controls, respectively.

**2.4.2 Catalase test**

An isolated single colony was picked from the plate using a wire loop and transferred onto a sterile glass slide. One drop of 30% hydrogen peroxide was added onto the culture on the glass slide and observed for any bubbles. Appearance of gas bubbles indicated that the isolate was catalase positive while absence of bubbles indicated a negative catalase test. *Pseudomonas flourescens* and *Lactobacillus plantarum* obtained from Mak Food Science culture collections, were included as the catalase positive and negative reactions, respectively.

**2.4.3 Oxidase test**

An oxidase identification strip (specs) was used to carry out the oxidase test. One end of the strip paper was rolled on the pure colonies from a 24 h plate culture and then left to stand for 1 min. A positive oxidase test was confirmed by observation of a color change of the strip from pink to purplish-black. The control samples included *Pseudomonas flourescens* and *Lactobacillus plantarum* for confirmation of oxidase positive and negative reactions, respectively.

**2.4.4 Extraction of genomic DNA**

The cetyl trimethyl ammonium bromide (CTAB) method described by Mugampoza et al. (2020) was used to extract genomic DNA. The bacterial isolate was cultured in nutrient broth for 24 h at 30-37ºC. Cells from 1 ml of the culture broth were centrifuged at 8000 rpm for 5 min. The cell pellet was re-suspended and washed twice by spinning at 8000 rpm at 4°C with 1 ml ice cold TE buffer (10 mM Tris-HCl, 1 mM EDTA, pH 7.5). Then, 30 µl of 20% sodium dodecyl sulphate (SDS) solution was added followed by 100 µl of CTAB solution. The mixture was incubated at 65°C for 10 min and 967 µl of 24:1 chloroform: isoamyl alcohol added. The resultant solution was mixed thoroughly and centrifuged at 13,000 rpm for 5 min. The upper phase supernatant (500 µl) was transferred into a clean micro centrifuge tube, 500 µl of ice-cold isopropanol added and gently mixed for 1 min to precipitate the DNA. The DNA was recovered by centrifugation at 13,000 rpm for 5 min at 4°C. Then, the DNA pellet was washed twice in 500 µl of 70% absolute ethanol, air dried for 30 min at room temperature, re-suspended in 100 µl TE buffer (10 mM Tris-HCl, 1 mM EDTA, pH 7.5) and stored at -25°C until analysis.

### 2.4.4.1 PCR amplification of the 16S rRNA gene

Primers V3F (CCTACGGGAGGCAGCAG) and V3R (ATTACCGCGGCTGCTGG) were used to amplify the variable V3 region of the 16S rRNA gene giving a PCR product of 200 bp. The reaction mixture of 50 µl contained 5 µl of 10X PCR buffer (10 mM Tris-HCl, 50 mM KCl, 1.5 mM MgCl, pH 8.3); 2.5 mM deoxynucleotide triphosphates; 0.2 pmol/µl (each) forward and reverse primers; 1.25 U of *Taq* DNA polymerase; and 1 µl of template DNA. The sample was amplified in a PCR thermocycler as follows: DNA denaturation for 5 min at 94°C followed by a touchdown PCR performed as follows: initial annealing temperature 66°C, and this was decreased 1°C every cycle for 10 cycles; finally, 20 cycles were performed at 56°C. The extension for each cycle was carried out at 72°C, 3 min, while the final extension was performed at 72°C, 10 min.

### 2.4.4.2 Gel Electrophoresis of PCR products

The PCR product (10 µl) was mixed with 2 µl of 6X loading dye and run on a 2% agarose gel containing 0.2 µg/ml ethidium bromide in 1X TAE buffer at 75 V, 2 h. A 100 bp DNA ladder was used as the molecular size marker. The gel was visualized on a UV trans illuminator (specs), and images recorded with Quantity One Gel Doc software (BioRad).

### 2.4.4.3 Sequencing and database search

Purification and sequencing of microbial isolates was done at Macrogen Meibergdreef 31, Amsterdam (Netherlands). The nucleotide sequences obtained were used to establish the closest known relatives of the isolates using the BLAST search tool and the NCBI gene bank.

**2.4.4.4 Confirmation of bacterial species**

Individual bacterial species were identified using the basic local alignment search tool (BLAST) programme

**3 Results and Discussion**

Our study aimed to identify microbial species in locally processed cheese using both conventional and molecular methods. Conventional techniques primarily identified LAB genera and yeasts, while PCR-based identification revealed a broader range of bacterial species.

Data obtained from conventional methods revealed that the cheese was dominated by lactic acid bacteria (LAB) (6.82±0.20 log CFU/g) and yeast & molds (2.54±0.05 log CFU/g) (table 2). None of the thermotolerant coliforms, *Staphylococcus* spp. and *Escherichia coli* were detected (LOD, 1 log CFU/g). The presumptive microbial genera (n=78 isolates) were; Gram positive, catalase negative and oxidase negative cocci (*Streptococcus*, *Enterococcus* or *Lactococcus* spp.), and Gram positive, catalase and oxidase negative rods (*Lactobacillus spp*.)

Conventional culturing and biochemical tests predominantly identified LAB genera, such as *Lactobacillus* and *Streptococcus*, as well as yeasts. Although these methods are useful, they have limitations in identifying less prevalent or non-culturable bacteria (Chao et al., 2003; Rantsiou & Cocolin, 2006)..

Using PCR and sequencing of the 16S rRNA gene, we detected additional bacterial species, including *Bacillus* spp. and *Clostridium* spp., which were not identified by conventional methods. The conventional methods' focus on culturable organisms and biochemical profiles might have led to an underrepresentation of the actual microbial diversity. PCR's higher sensitivity and broader detection capabilities allowed us to identify species that were either not present in significant quantities to be detected by conventional methods or were not cultivable on the media used.

The discrepancy highlights the limitations of relying solely on conventional methods for microbial identification. PCR-based techniques provide a more comprehensive overview of microbial diversity, revealing species that are often missed by traditional methods.

TABLE 2: Enumeration of microorganisms in ripened cheese produced locally and internationally

| Samples | TPC | Total Coliforms | Total LAB | Lactococcus | Thermo-Tolerant coliforms | Yeasts |
|---|---|---|---|---|---|---|
| **Cheddar** | 6.87±0.04 | 1.90±0.01 | 6.72±0.06 | 6.83±0.05 | ND | 2.55±0.23 |
| **Jack** | 7.91±0.03 | ND | 7.51±0.22 | 7.60±0.25 | ND | 2.31±0.26 |
| **Gouda local** | 6.42±0.30 | 2.0+0.01 | 5.75±0.09 | 6.21±0.30 | ND | 2.48±0.16 |
| **Permisan** | 6.12±0.51 | ND | 5.83±0.65 | 5.81±0.38 | ND | 2.64±0.19 |
| **Imported gouda 1** | 7.19±0.12 | ND | 6.88±0.04 | 6.90±0.03 | ND | 2.88±0.04 |
| **Imported Gouda 2** | 4.15±0.21 | ND | 4.69±0.13 | 4.48±0.02 | ND | ND |
| **Imported Gouda 3** | 8.91±0.05 | ND | 7.82±0.05 | 8.06±0.08 | ND | 2.74±0.06 |

Values are expressed as log cfu/g

Table 3: Presumptive Identification of Bacteria isolated from locally processed and imported cheese based on conventional microbiology methods

| Samples | Group | Catalase | Oxidase | Gram Stain | Colony morphology | Cell shape | Number (%) of Isolates | Presumptive identification |
|---|---|---|---|---|---|---|---|---|
| **Cheddar, Jack Permisan, Gouda-local Imported Gouda 1,2,3** | 1 | - | - | + | Cream, Round | Cocci | 192 (89.3) | Streptococcus, Enterococcus Lactococcus |
| **Cheddar, Jack and Gouda-local** | 2 | - | - | + | Cream, Round | Rods | 23 (10.7) | Lactobacillus, Clostridium |

Table 4. Results of Blast sequence analysis obtained from PCR amplification of 16S rDNA of bacteria isolated from selected ripened cheese varieties produced in Uganda

| Group | Gene accession | Close relative of | Number of isolates | Proportion (%?) | E-value | %ID |
|---|---|---|---|---|---|---|
| 1 | MK208576.1 | Bacillus haynesii | 1 | 0.9 | 5e-08 | 89 |
| 2 | KY615355.1 | Bacillus mycoides | 1 | 0.9 | 3e-72 | 99 |
| 3 | MH628186.1 | Bacillus sp. | 1 | 0.9 | 3e-71 | 98 |
| 4 | KR812508.1 | Bacterium | 4 | 3.5 | 6e-30 | 96 |
| 5 | KM457447.1 | Enterococcus hirae | 1 | 0.9 | 8e-74 | 98 |
| 6 | AY659871.1 | Enterococcus sp. | 1 | 0.9 | 6e-64 | 96 |
| 7 | GU429394.1 | Lacticaseibacillus rhamnosus | 10 | 8.8 | 5e-75 | 100 |
| 8 | MF357249.1 | Lactiplantibacillus paraplantarum | 1 | 0.9 | 4e-61 | 96 |
| 9 | CP054259.1 | Lactiplantibacillus plantarum | 2 | 1.8 | 6e-68 | 97 |
| 10 | MG722901.1 | Lactobacillus brevis | 1 | 0.9 | 8e-48 | 93 |
| 11 | KM921935.1 | Lactobacillus casei | 2 | 1.8 | 5e-75 | 100 |
| 12 | KX951732.1 | Lactobacillus paracasei | 1 | 0.9 | 1e-71 | 98 |
| 13 | KJ775808.1 | Lactobacillus plantarum | 3 | 2.6 | 2e-73 | 99 |
| 14 | MT780865.1 | Lysinibacillus sp. | 1 | 0.9 | 3e-10 | 90 |
| 15 | MT045929.1 | Pediococcus acidilactici | 7 | 6.1 | 1e-70 | 98 |
| 16 | JN039352.1 | Pediococcus pentosaceus | 26 | 23.2 | 3e-72 | 98 |
| 17 | KC191582.1 | Pseudomonas fluorescens | 1 | 0.9 | 2e-53 | 92 |

| 18 | EU482921.1 | Pseudomonas sp. | 6 | 5.3 | 1e-73 | 99 |
| 19 | KX648542.1 | Staphylococcus epidermidis | 1 | 0.9 | 3e-71 | 99 |
| 20 | KM819142.1 | Uncultured Bacillus sp. | 2 | 1.8 | 4e-67 | 97 |
| 21 | MH229146.1 | Uncultured bacterium | 18 | 15.8 | 5e-08 | 100 |
| 22 | KX672804.1 | Uncultured Lactobacillus sp. | 3 | 2.6 | 2e-22 | 90 |
| 23 | HQ774830.1 | Uncultured microorganism | 2 | 1.8 | 1e-23 | 93 |
| 24 | JF427767.1 | Uncultured Pediococcus sp. | 15 | 13.2 | 1e-74 | 99 |
| 25 | EU339600.1 | Uncultured soil bacterium | 1 | 0.9 | 4e-14 | 92 |
| 26 | AB250247.1 | Uncultured Streptococcus sp. | 1 | 0.9 | 1e-43 | 89 |
|  |  | Total | 114 | 100 |  |  |

Using the blast program against the NCBI GenBank database, we set an E-value threshold of 1e-5 and a minimum identity of 90%. We performed a BLAST search on our microbial genome sequence to identify potential homologs and determine the species classification. BLAST analysis revealed that *Pediococcus pentosaceus* (accession number *JN039352.1*), showing a 98% sequence identity and an E-value of 3e-72 dominated in all cheese samples (23.7%, of the total 114 isolates). Other significant hits included *Pediococcus acidilactici* (13.2%), *Lacticaseibacillus rhamnosus* (8.8%) indicating close evolutionary relationships as shown in Table 4.

Lactic acid bacteria are well known for their roles in cheese ripening and flavor development. Lactobacillus casei, for instance, contributes to the breakdown of proteins and fats, enhancing the cheese's texture and taste ((Qi et al., 2021). Our findings were consistent with previous studies, which have also identified these species in ripened cheeses (Montel et al., 2014; McSweeney et al., 2004). LAB enhance quality properties of fermented foods. Several studies have indicated isolation of different species of LAB from cheese samples (Qi et al., 2021) as cheese is nutrient rich- easy digestible, palatable.

*Pediococcus pentosaceus* has been identified as a key microorganism responsible for flavor development and accelerated fermentation in Manura cheese, as reported by Gerasi et al. (2010). Ilaria et al. (2015) found several strains of P. pentosaceus in traditional Italian Marga cheese, indicating its natural presence or potential for use in cheese production.

P. *pentosaceus* enhances cheese flavor mainly through its fermentation abilities. It produces metabolites that contribute to flavor formation and cheese ripening, improving sensory qualities. This bacterium is particularly effective at breaking down proteins and fats, releasing flavor compounds that enhance the taste and texture of cheese (Qi et al., 2021). Its proteolytic activity results in smaller peptides and amino acids, essential for flavor development, producing compounds such as amino acids, peptides, and sulfur compounds (Gobbetti et al., 2002).

Additionally, McSweeney and Sousa (2000) noted that P. *pentosaceus* facilitates lipolysis, releasing free fatty acids, which are vital precursors for volatile compounds that contribute to the aroma of cheese. This bacterium also exhibits high salt tolerance and can produce bioactive compounds like gamma-aminobutyric acid (GABA), which benefit both flavor enhancement and health (Ugras et al., 2024; Pervez et al., 2006). Its resilience during cheese processing ensures its effectiveness in improving cheese quality.

Furthermore, P. *pentosaceus* produces lactic acid during fermentation, lowering the cheese's pH. This acidic environment is conducive to developing desirable flavors and textures (Fernández et al., 2015).

## 4 Conclusion

Microbial analysis of different ripened cheese varieties reveals a rich and diverse microbial ecosystem that plays a crucial role in the ripening process, contributing to the unique sensory properties of each cheese. The results highlight that while starter cultures are used to initiate fermentation, non-starter LAB and natural microflora present in the environment contribute significantly to the complexity of the final product. This demonstrates the value of integrating molecular techniques with conventional methods to obtain a more accurate picture of microbial communities.

Finally, the findings suggest that understanding the microbial profiles of ripened cheeses can inform the development of tailored starter cultures, improving the consistency, quality, and safety of cheese production. Future studies should consider using both approaches to capture the full spectrum of microbial diversity.

## Acknowledgements

This work was funded by Kyambogo University and special thanks to the staff of Bioresearch laboratory of National Agriculture Research Organisation (NARO) where PCR samples were prepared.